\documentclass{llncs}

\usepackage{graphicx}
\usepackage{color}
\usepackage{amsmath}
\usepackage[dvipsnames]{xcolor}
\usepackage[normalem]{ulem}
\usepackage{tabularx}
\usepackage{booktabs}
\usepackage{calc}
\usepackage{overpic}
\usepackage{colortbl}

\definecolor{C1}{RGB}{231,69,51}
\definecolor{C2}{RGB}{26,198,83}
\definecolor{C3}{RGB}{56,110,165}
\definecolor{C4}{RGB}{11,84,1}
\definecolor{C5}{RGB}{253,148,7}
\definecolor{C6}{RGB}{0,0,109}
\definecolor{C7}{RGB}{230,28,121}
\definecolor{C8}{RGB}{118,13,21}

\newcommand{\Cone}[1]{\textcolor{C1}{#1}}
\newcommand{\Ctwo}[1]{\textcolor{C2}{#1}}
\newcommand{\Cthree}[1]{\textcolor{C3}{#1}}
\newcommand{\Cfour}[1]{\textcolor{C4}{#1}}
\newcommand{\Cfive}[1]{\textcolor{C5}{#1}}
\newcommand{\Csix}[1]{\textcolor{C6}{#1}}
\newcommand{\Cseven}[1]{\textcolor{C7}{#1}}
\newcommand{\Ceight}[1]{\textcolor{C8}{#1}}

\begin{document}

\title{Visual Similarity Perception of Directed Acyclic Graphs: A Study on Influencing Factors}
\titlerunning{Similarity Perception of Directed Acyclic Graphs}  
%
\author{K. Ballweg\inst{1} \and M. Pohl\inst{2} \and G. Wallner\inst{2} \and T. von Landesberger\inst{1}}
\authorrunning{Ballweg et al.} 
%
\tocauthor{Kathrin Ballweg, Margit Pohl, G\"unter Wallner, and Tatiana von Landesberger}
\institute{Technische Universit\"at Darmstadt, Darmstadt, Germany,\\
\email{\{kathrin.ballweg,tatiana.von.landesberger\}@gris.tu-darmstadt.de}
\and
Vienna University of Technology, Vienna, Austria,\\
\email{margit@igw.tuwien.ac.at, guenter.wallner@tuwien.ac.at}
}

\maketitle             

\begin{abstract}
While visual comparison of directed acyclic graphs (DAGs) is commonly encountered in various disciplines (e.g., finance, biology), knowledge about humans' perception of graph similarity is currently quite limited. By graph similarity perception we mean how humans perceive commonalities and differences in graphs and herewith come to a similarity judgment. As a step toward filling this gap the study reported in this paper strives to identify factors which influence the similarity perception of DAGs. In particular, we conducted a card-sorting study employing a qualitative and quantitative analysis approach to identify 1) groups of DAGs that are perceived as similar by the participants and 2) the reasons behind their choice of groups. Our results suggest that similarity is mainly influenced by the number of levels, the number of nodes on a level, and the overall shape of the graph.
	    
\keywords{Graphs, Perception, Similarity, Comparison, Visualization}
\end{abstract}

\section{Introduction} 
	\label{sec:introduction}
	
The visual comparison of directed acyclic graphs (DAGs) is a task encountered in various disciplines, e.g., in finance, biology, natural language processing, or social network analysis. The task is strongly influenced by the human perception of similarity since comparison builds upon making similarity judgments. In spite of the numerous occurrences of this task and recent papers surveying visual graph comparison techniques and visualizations~\cite{beck2014state,gleicher2011visual}, knowledge about the human perception of graph similarity -- especially for DAGs -- is quite limited.

Only a few investigations address the comparison of graphs. Gleicher et al.~\cite{gleicher2011visual} identify the basic types of techniques for visual comparison (juxtaposition, superposition, and explicit encoding). Tominski et al.~\cite{Tominski:2012} explicitly deal with the comparison of large node-link diagrams in superposition. They argue that interaction is essential for this process.
Some interesting insights can be gained from the literature on dynamic graphs showing the evolution of node-link diagrams over time. The survey of Beck et al.~\cite{beck2014state} about visualizations of dynamic graphs provides an overview of visualization options transferable to general visual graph comparison since dynamic graph visualization has an inherent comparison component. Others discuss the extension of these visualizations and techniques with features like highlighting of commonalities and differences or the effectiveness of difference maps~\cite{archambault2009structural,Archambault:2011,bach2014graphdiaries,bremm2011interactive,holten2008visual}. However, none of these papers deal with the issue of similarity perception within the context of graph comparison.
There is also a large amount of research on graph readability. This research is partially relevant since the DAGs need to be read in order to compare them. Examples include studies on edge crossings and mental map perseverance (e.g.,~\cite{kobourov2014crossings,purchase2001graph,purchaseaesthetics,purchase2002metrics}. These aspects are, however, not in the focus of our attention.
The research investigating the comparison of visualizations in general is also interesting. Pandey et al.~\cite{pandey2016towards} conducted an experiment to study the similarity perception of scatterplots. So, their work inspired our methodology.

To the best of our knowledge, there is no research focusing on how humans perceive the similarity of DAGs. We are especially interested in the \emph{factors which influence the perception of similarity} (possibly, number of nodes/edges, edge crossings, etc.). We deem the knowledge about the influencing factors important for the generation of future actionable guidelines for comparative visualizations. Towards this end, we conducted a study with small, unlabeled synthetic DAGs and used card sorting as our methodology. We decided for these DAGs in order to be able to keep the number of to be tested factors manageable. However, because of our systematic procedure the study scope can be easily extended in the future.
The DAGs are represented as node-link diagrams. We address two research questions: (1) \emph{Which groups do the participants form?}, and (2) \emph{Which factors did the participants consider to judge the similarity?}


Our results indicate that similarity is mainly influenced by the number of levels and the number of nodes on a level as well as the overall shape. We provide additional material - i.a. our study material, our collected data and our analysis results here:\\ \href{http://www.gris.tu-darmstadt.de/research/vissearch/projects/DAGSimilarityPerception/index.html}{\underline{\small{www.gris.tu-darmstadt.de/research/vissearch/projects/DAGSimilarityPerception}}}.

\section{Related Work} 
	\label{sec:related_work}

While there exists an extensive body of research in perceptual psychology and pattern recognition on similarity judgments and dissimilarity measures (cf. e.g., \cite{Robert:2005,Pekalska:2005} for an overview), we will concentrate on work dealing with graphs and other types of plots. 

Starting with graph visualization techniques and visual comparison techniques there exist several recent surveys of these areas (e.g.,~\cite{beck2014state,gleicher2011visual,hadlaksurvey,vehlow15state,von2011visual}). The basic techniques, that is, juxtaposition, superposition, and explicit encoding -- following Gleicher et al.'s~\cite{gleicher2011visual} classification -- are sometimes enriched by emphasizing the correspondences between graphs~\cite{vislink,holten2008visual}, e.g., by highlighting similar parts~\cite{bach2014graphdiaries,bremm2011interactive,holten2008visual}, or by emphasizing differences by collapsing the identical parts~\cite{archambault2009structural}. The enrichment, that is, emphasizing the commonalities or differences, usually relies on a similarity function respecting specific criteria. In this respect, Gao et al.~\cite{Gao:2010} provide an overview of research done on graph edit distances, a mathematical way to measure the similarity between pairwise graphs. However, it is still unknown whether the criteria on which existing similarity functions are based correspond to the criteria used by humans when visually comparing two or more node-link diagrams. Tominski et al.~\cite{Tominski:2012} proposed interaction techniques which aid users in doing comparison tasks and which were inspired by the real-world behavior of people when comparing information printed on paper. Getting a better understanding of the perceived differences and commonalities is likely to result in better visualization and interaction techniques.

Moreover, the existing body of work dealing with perceptual and cognitive aspects focuses primarily on the readability of single graphs. Several factors, including graph aesthetics (edge crossings~\cite{kobourov2014crossings,purchase2002metrics,purchase2007important}, layout~\cite{mcgrath1997effect,dwyer2009comparison,Kieffer2016,mcgee2012empirical,novick2006importance}, graph design~\cite{Holten:2009,tennekes2014tree}, and graph semantics or content knowledge~\cite{purchase2001graph,novick2006importance,korner2005concepts}) have been identified to be important for graph readability. Huang et al.~\cite{Huang:2006} -- concerned with sociograms --  note that good readability is not enough to effectively communicate network structures, emphasizing that the spatial arrangement of the nodes also influences viewers in perceiving the structure of social networks. 

While perceptual aspects of single graphs have been thoroughly investigated, literature dealing with perceptual aspects when comparing node-link visualizations is considerably more scarce. Notable papers in this space are the work of Bach et al.~\cite{bach2014graphdiaries} and Ghani et al.~\cite{Ghani:2012} who are both concerned with dynamic graphs (cf. Beck et al.~\cite{beck2014state} for an overview). Noteworthy is also the work of Archambault et al.~\cite{Archambault:2011} who evaluated the effectiveness of difference maps which show changes between time slices of dynamic graphs. While we are not necessarily concerned with dynamic graphs these works are nonetheless relevant in our context as dynamic graphs are often analyzed by using discrete time-slices. In our own previous work~\cite{Landesberger:2017} we provided an overview of methodological challenges when dealing with the investigation of graph comparison and described a first preliminary study targeted towards identifying factors which influence the recognition of graph differences in very small star-shaped node-link diagrams. The work presented in this paper can be viewed as a continuation of these efforts. 

Looking beyond the perception of node-link diagrams literature is currently also quite limited when it comes to similarity perception of other types of plots, a sentiment shared by Pandey et al.~\cite{pandey2016towards} who investigated how human observers judge the similarity of scatterplots. Our quantitative analysis as presented in this paper is partly based on the methodology put forward by Pandey et al.\;. Fuchs et al.~\cite{Fuchs:2014} looked into how contours affect the recognition of data similarity in star glyphs. Likewise, Klippel et al.~\cite{Klippel:2009} investigated the similarity judgments of star glyphs using a methodology comparable to the one used by Pandey et al.~\cite{pandey2016towards} and us: Participants were shown various plots which they then had to group according to their perceived similarity.

\section{Study Methodology}
	\label{sec:study_methodology}

In this section, we present our study methodology. As noted above, Pandey et al.'s~\cite{pandey2016towards} work about the human similarity perception of large sets of scatterplots strongly inspired our basic methodology, since we share the research questions for different data types. Furthermore, Pandey et al. substantiate that the methodological principle of card sorting produces valuable results for this type of research questions. For advantages, drawbacks, and the suitability of card sorting for research questions like ours see Section~\ref{sub:study_procedure}. 

\subsection{Research Questions} 
	\label{sub:research_questions}

Our superordinate research question (RQ) is: \emph{What factors influence human similarity perception of DAGs?} We firstly have to know the factors influencing the similarity judgment. Once we know the influencing factors, we can, for instance, research the specific degree of influence of a single factor as well as the interplay between the factors. In order to analyze our superordinate RQ, we formulate two subordinate ones: 

\begin{itemize}
	\item RQ1: \emph{Which groups do the participants form?}
	\item RQ2: \emph{Which factors did the participants consider to judge the similarity?}
\end{itemize}

\subsection{Dataset} 
	\label{sub:dataset}

Creating an appropriate study dataset is challenging due to the large number of possible variations~\cite{Landesberger:2017}. Therefore, we were forced to limit the number of DAGs. Our object of study were 69 small (6-9 nodes), unlabeled, synthetic DAGs visualized as node-link diagrams. In the following we will use the term DAG to also refer to its embedding. When creating the DAGs we considered \emph{known factors influencing graph readability} (e.g., edge crossings) and \emph{characteristics of DAGs from real-world datasets} (e.g., a node may be the child of more than one parent node). We decided to have synthetic and small DAGs in order to be able to keep the number of to be tested factors manageable and to evaluate them systematically.
Because of our study and data creation methodology it is easily feasible to extend our results with further studies considering further factors. Especially since knowledge about human graph similarity perception is currently quite limited, we consider the manageability of the problem as crucial. The size of our DAGs is also realistic. They are comparable to cascades in finance and biology (e.g.,~\cite{von2015visual,lenz2014visual}) and directed acyclic word graphs~\cite{thornley2013using}.\\ 
We deem \emph{factors of graph readability} as potentially important for visual graph comparison since in order to be able to visually compare DAGs it is necessary to read them. We consider \emph{properties of real DAGs} as important for our studies since they influence the visual appearance of the DAGs. More importantly, considering properties of real DAGs strengthens the \emph{realism} of our synthetic data and, consequently, the \emph{transferability} of our results to real world use cases.

\begin{figure}[t]
  \centering
  \includegraphics[width=\textwidth]{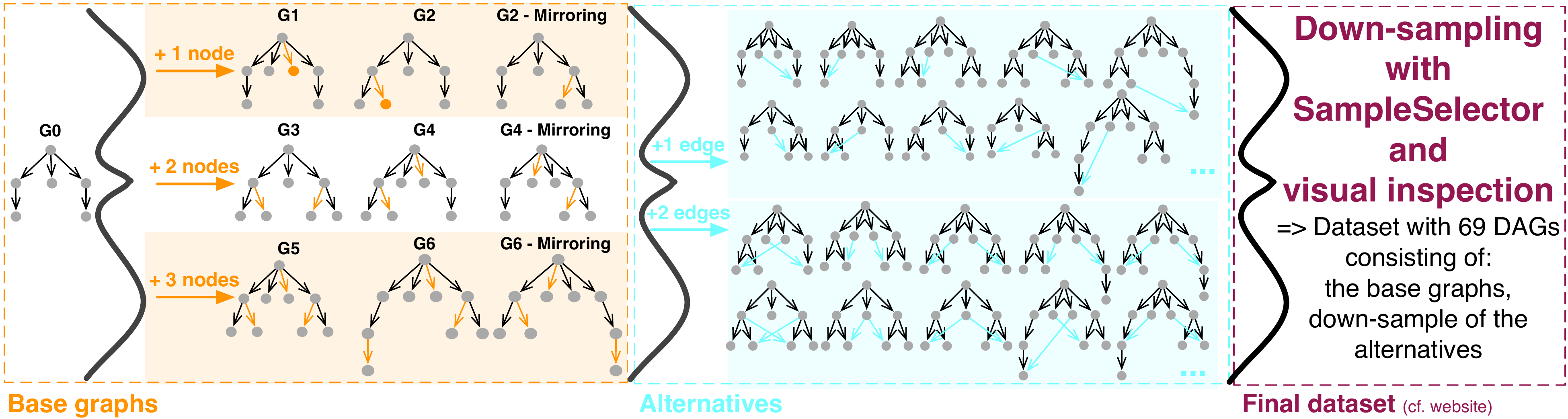}
  \caption{Dataset creation: (I) Base graph creation by adding 1, 2, and 3 nodes to $G0$ ensuring that the added node is placed at the \emph{inner as well as the outer areas of $G0$}, (II) creation of all possible alternatives by adding one and two edges to the base graphs, (III) down-sampling of the alternatives considering the factors described in Section~\ref{sub:dataset}.}
  \label{fig:pictures_DAG_CreationProcessEPS}	
\end{figure}

To create our dataset, we started with $G0$, as depicted in Figure~\ref{fig:pictures_DAG_CreationProcessEPS}. $G0$ is symmetric since it is easier to break symmetry than to introduce symmetry. We herewith cover \emph{symmetric} and \emph{asymmetric} DAGs on the basis of $G0$. This is important since humans are prone to symmetry~\cite{WelchKobourov_symmetry}. $G0$ is single-rooted since this is typical for various real-world DAG datasets; e.g., cascades. To test \emph{node} and \emph{edge changes (addition of node(s) and edge(s))} we had a two stage DAG creation process. First, we created the base graphs $G1$ - $G6$ and their reflections by adding one, two, and three nodes. We ensured that the addition of the node(s) is done in the \emph{inner as well as the outer areas of $G0$} (cf. Figure~\ref{fig:pictures_DAG_CreationProcessEPS} - Base graphs). Secondly, we created all possible DAGs resulting from adding one and two edges (cf. Figure~\ref{fig:pictures_DAG_CreationProcessEPS} - Alternatives) using our custom-made \emph{GraphCreator} -- a tool to create all possible DAGs resulting from a specific change of a DAG, e.g., adding one edge to $G1$ - $G6$ and their reflections. Herewith, we ensure that we have the maximal possible variation from which we then sample our study dataset.\\
Down-sampling is necessary since the visual comparison of DAGs is a quite cognitive demanding task for the participants. For the down-sampling we considered the following factors: \emph{edge crossing, visual layout, more than one parent node has the same child node}, and \emph{long connections -- typically across more than one level}. We considered \emph{edge crossing} since it is a prominent factor in graph readability~\cite{kobourov2014crossings,purchase2002metrics} for which reason we assume that it also plays a role in visual graph comparison. Furthermore, we considered the \emph{visual layout} -- another known factor from graph readability. The visual layout of DAGs does not contain any analytically relevant information about the DAGs' data structure and properties. However, it still has a significant impact on graph readability which is why we deem it important to test its influence on visual graph comparison~\cite{dwyer2009comparison,Huang:2006}. We test this factor by horizontally \emph{reflecting} $G2$, $G4$, and $G6$ (cf. Figure~\ref{fig:pictures_DAG_CreationProcessEPS}). The ability to test the impact of \emph{isomorphism} is a beneficial byproduct of this decision. We decided for a traditional hierarchical node-link diagram layout (\includegraphics[height=0.4cm]{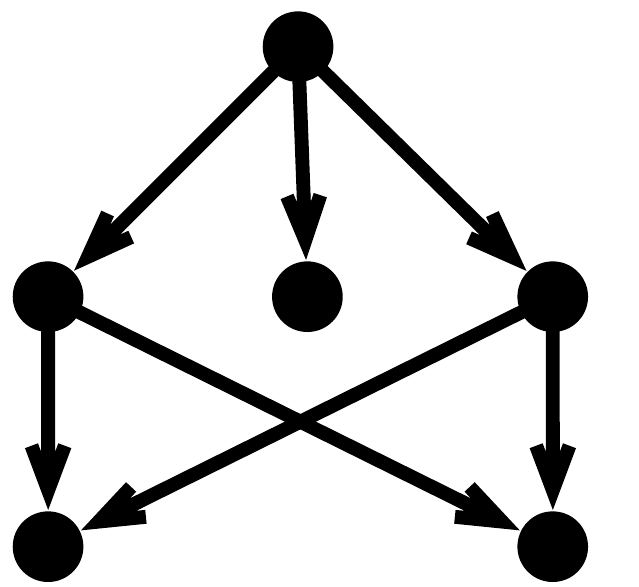}) with the root placed on top since Burch et al. found that this layout type outperforms other types such as upward layouts. In order to avoid confounding effects by destroying the mental map we did not optimize the layout after a DAG change; e.g., resolving an avoidable edge crossing. By visually inspecting DAGs from real-world datasets we found that two frequently occurring properties are: \emph{more than one parent node has the same child node} (\includegraphics[height=0.4cm]{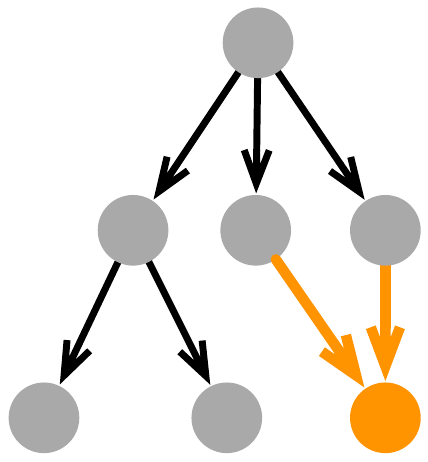} -- e.g., cascades in finance and biology), \emph{long connections -- typically across more than one level} (\includegraphics[height=0.4cm]{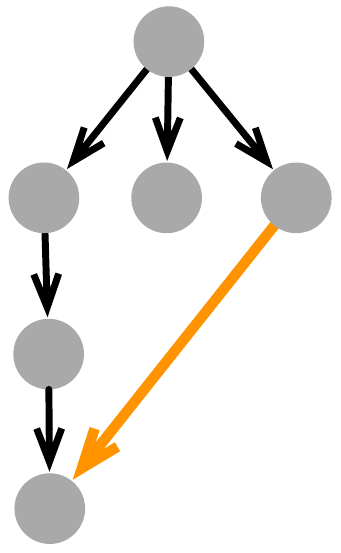} -- e.g., directed acyclic word graphs in natural language processing). Consequently, we considered them as factors for our final dataset. We did the down-sampling under the constraint of preserving the systematic variation (cf. Figure~\ref{fig:pictures_DAG_CreationProcessEPS} -- Final dataset) and by ensuring that changes take place at the \emph{inner as well as the outer areas of the respective base graph}. Our dataset is available here: \href{http://www.gris.tu-darmstadt.de/research/vissearch/projects/DAGSimilarityPerception/index.html}{\underline{website}}.

\subsection{Participants} \label{sub:participants} We recruited 20 volunteers (13 male, 7 female, between 20 and 60 years). We had no prerequisite of having experience with DAGs. This way our results are not limited to experienced users. In our opinion, it is more likely that experienced users know which factors really bear relevant information for the comparison task whereas for inexperienced users misconceptions are more likely. We are convinced that if we want to understand the human similarity perception and as a consequence improve comparative visualizations, we need a varying range of expertise with DAGs. Our participants had a diverse educational level (vocational training, undergraduate, graduate, post-graduate) and came from various disciplines. Two of our participants had basic knowledge in information visualization and five had advanced knowledge. In spite of this, the participants' experience with DAGs varied vastly. 

\subsection{Study Procedure} 
	\label{sub:study_procedure}
Every participant was welcomed and the experimenter handed over the study material and explained the task. Each session took approximately one hour.

\paragraph{Task.} 
	\label{par:task}

We asked the participants to group 69 DAGs with respect to their perceived similarity -- multiple occurrences of a single DAG in different groups were allowed. Furthermore, we asked them to tag each group with the factors they used to build them. Finally, participants had to judge the easiness of forming the respective group (\emph{``How difficult or easy was it for you to create this group?''}) and their confidence in the group's consistency (\emph{``How doubtful or confident are you about the consistency of the DAGs in the group, i.e., would you create the same group again if you did this task again?''}). The questions regarding easiness and confidence were to judge on a five-point Likert scale (``1 = very difficult/doubtful, 2 = difficult/doubtful, 3 = neutral, 4 = easy/confident, 5 = very easy/confident''). The formed groups provide the data needed to answer \emph{RQ1} while the participants' group tags provide the data to answer \emph{RQ2}. For the task formulation we kept the one from Pandey et al.~\cite{pandey2016towards} since it captured exactly what we wanted to ask our participants. Moreover, the formulation was already pretested and successful in Pandey et al.'s study.

\paragraph{Card Sorting Methodology.} 
	\label{par:card_sorting_methodology}

Card sorting is a well-known methodology in psychology and human-computer interaction for externalizing mental models humans have about the environment they live in. Wood and Wood~\cite{Wood:2008} define card sorting as follows: \emph{As the name implies, the method originally consisted of researchers writing labels representing concepts (either abstract or concrete) on cards, and then asking participants to sort (categorize) the cards into piles that were similar in some ways.}  Humans group objects according to their perceived similarity into different categories. In this way, card sorting helps to uncover the structure of mental models. There are different methods to conduct card sorting. Researchers generally distinguish between open vs. closed sorting tasks and between paper-based and computer-supported card sorting~\cite{Greve:2014}. In closed card sorting, participants have to sort the cards according to a given scheme, in open card sorting, the participants develop it themselves. The procedures for card sorting tasks sometimes differ considerably. Sometimes, the cards that have been assigned to a category are placed in a pile~\cite{Wood:2008}, so that participants do not shuffle them around on a canvas. Especially in computerized card sorting, it is often not possible to see all cards from which to choose at the same time~\cite{Chaparro:2008,pandey2016towards} which forces the study participants to compare the cards in memory. We used an open, paper-based card sorting since literature indicates that the paper-based approach yields more consistent results than the computerized one~\cite{Greve:2014}. 

\paragraph{Study Setup and Materials.}  
	\label{par:study_setup_and_materials}

We used an empty meeting room with good lighting for conducting the study. The participant got the task sheet, the data sheet, sheets for building the groups, and sheets for tagging each group with the group building factors as well as for judging the easiness and the confidence. The task sheet contained the afore explained task. The data sheet consisted of the 69 randomly positioned DAGs. We decided to present our dataset on paper, so that the participants could see all data items at the same time. The order of the data items was kept the same for all participants to exclude order as a possible confounding variable. The participants had to write down the DAGs' IDs which belong to a group and give each group a unique identifier. Furthermore, they had to write down the tags as well as their easiness and confidence judgment together with the unique group identifier. The material is available on our website.

\section{Analysis and Results} 
	\label{sec:analysis_and_results}

To analyze the collected data with respect to our research questions we used a mixed-approach involving a quantitative (RQ1, RQ2) and a qualitative (RQ2) analysis. The qualitative analysis provided the factors the participants tagged their formed groups with. The quantitative analysis resulted in the perceptual consensus over all participants as well as it served as a check of the participants' self-reported factors extracted in the qualitative analysis. 

\subsection{Quantitative Analysis (RQ1)}\label{subsec:quantitative_analysis}

We did a perceptual consensus calculation over all participants with complete data ($16$), i.e., participants who assigned each DAG to at least one group. 
The perceptual consensus of the perceived similarity served as a basis to find out (1) whether the similarity perception of humans is consistent across individual people and (2) whether it is objectifiable with graph theoretical or visual properties. 

To gain insights into the consistency and objectifiability of the similarity perception and to mitigate the potential bias -- saying one thing and doing another -- of self-reporting questions such as tagging we analyzed the perceived similarity consensus regarding which of the known graph theoretical and visual properties explain the clusters best. The mitigation potential resides in the perceived similarity consensus encapsulating what the participants really did.

The consensus easiness and confidence scores for each cluster provide information about the similarity consensus' perceived solidness and robustness. A high average easiness score means that the grouping is solid, thus, due to an easy assignment, it is less likely that a participant assigned a DAG randomly. The average confidence score reflects the participants' opinion whether they would form the same group again. A high score means that the grouping is robust since it is highly probable that it would look similar if the task were repeated.

\paragraph{Analysis.} To build the perceptual consensus for the participants' similarity judgments, we calculated a pairwise perceptual distance between each pair of DAGs, based on the number of occurrences of each DAG pair in the same group and on the number of individual occurrences (for details cf.~\cite{pandey2016towards}). The perceptual distance calculation resulted in a $69\times69$ perceptual distance matrix (PDM). Like Pandey et al.~\cite{pandey2016towards} we did a hierarchical clustering, in our case with average linkage. We evaluated the correct number of clusters with the mean/median of number of groups and with the gap statistic~\cite{Tibshirani:2001}. The mean/median indicate the average number of participant-built groups and thus served as a reasonable estimator for the number of clusters. The gap statistic respects, like the individual groupings and similarity per se, the cluster similarity which made it to a further reasonable estimator. The hierarchical clustering result is the consensus grouping of all DAGs based on the similarity consensus contained in the PDM. 

For the clusters' property analysis we determined various properties for each graph. Based on this we determined the dominating properties of the clusters as well as the cluster separating properties. Examples of the employed properties are: depth, visual symmetry, visual leaning, edge crossing -- number and existence, edge length, number of nodes on a specific level, and the existence and the number of nodes having more than on parent node.

For the consensus of the easiness and confidence score we calculated an easiness and confidence value for each plot on the basis of the assumption that each plot inherits the easiness and confidence score of the participant-built groups it belongs to. Then we calculated an average easiness and confidence score for the hierarchical clustering result. For a detailed explanation please refer to~\cite{pandey2016towards}.

\begin{figure}[t]
  \centering
    \includegraphics[width=\textwidth]{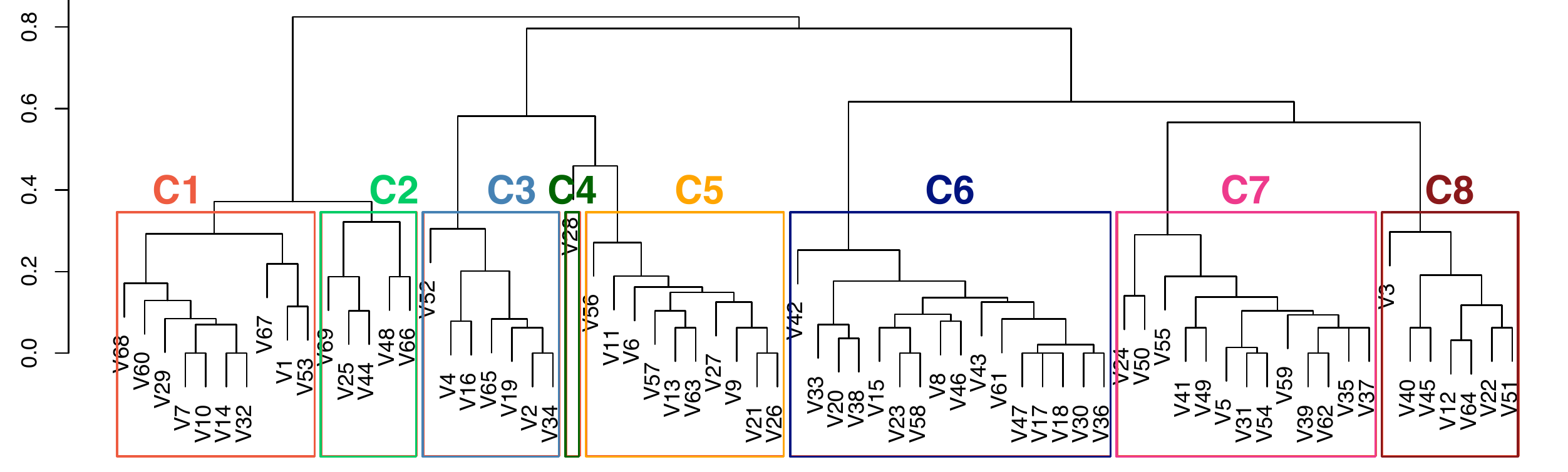}
  \caption{Dendrogram resulting from hierarchical clustering with average linkage. The resulting eight clusters (C1--C8) are marked using colored boxes.} 
  \label{fig:pictures_MDS_Dendrogramm_DAGs}
\end{figure}

\begin{figure}[t]
  \centering
    \includegraphics[width=\textwidth]{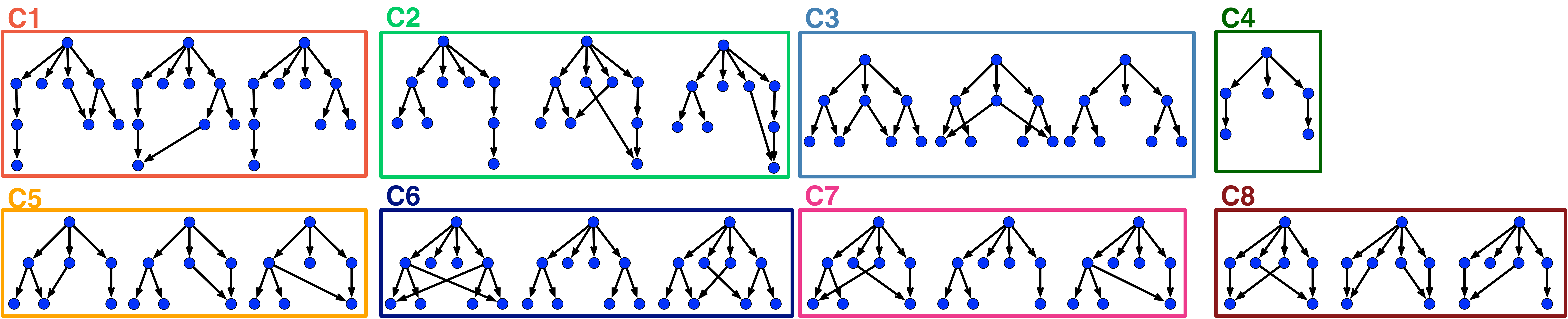}
  \caption{Excerpts of the hierarchical clusters (cf. website complete clusters).}
  \label{fig:pictures_ClusterExcerpts}
\end{figure}

\paragraph{Results.} The gap statistic indicated that the data supports eight clusters. Both the mean and median of the number of built groups supported the indicated eight clusters ($mean=7.6, median=8.0, STD=2.6$). As all three were similar, we decided to cut the tree into eight clusters. Figure~\ref{fig:pictures_MDS_Dendrogramm_DAGs} shows the resulting dendrogram and the resulting clusters -- marked using colored boxes. Excerpts of the hierarchical clusters are shown in Figure~\ref{fig:pictures_ClusterExcerpts}. The entire clusters can be found on our website. The easiness and confidence scores of all hierarchical clusters are around $4.0$ (cf.~Table~\ref{table:clustersAnalysis}). This means that the participants on average found their groups \emph{easy} to build and were \emph{confident} they would look similar if they repeated the task. Consequently, this results in a good solidness and robustness of the consensus grouping.

The properties which distinguish the clusters best are the \emph{depth} of the DAGs, the \emph{number of nodes on a specific level} of the DAGs, and the \emph{visual leaning} of the DAGs. Table~\ref{table:clustersAnalysis} summarizes the properties of the clusters. Clusters \Cone{C1} and \Ctwo{C2} are identical in depth and number of nodes on each of their four levels. However, they are separated by the leaning. While the DAGs of \Cone{C1} are left-skewed, those of \Ctwo{C2} are right-skewed. The leaning separating the clusters \Cone{C1} and \Ctwo{C2} suggests that not the reflection of $G6$ (cf. Figure~\ref{fig:pictures_DAG_CreationProcessEPS}) itself was apparent to the participants but rather a property which changed -- the leaning (cf. Section~\ref{sub:dataset}).\\
Clusters \Cthree{C3}, \Cfour{C4}, and \Cfive{C5} have identical depth (3) as well as three nodes on the second level. The number of nodes on the third level separates these clusters. The depth separates the clusters \Cthree{C3}, \Cfour{C4}, \Cfive{C5} from \Cone{C1}, \Ctwo{C2}. Cluster \Cfive{C5} clearly shows that neither the reflection of $G2$ (cf. Figure~\ref{fig:pictures_DAG_CreationProcessEPS}) itself nor a changed property mattered. It seems that the pure number of nodes dominates significantly over, e.g., node position (2 left, 1 right vs. reflected: 1 left, 2 right).\\
Clusters \Csix{C6}, \Cseven{C7}, and \Ceight{C8} have identical depth (3) and four nodes on the second level. The number of nodes on the third level separates them. The number of nodes on the second level separates \Csix{C6}, \Cseven{C7}, \Ceight{C8} from \Cthree{C3}, \Cfour{C4}, \Cfive{C5}. \Csix{C6}, \Cseven{C7}, \Ceight{C8} and \Cone{C1}, \Ctwo{C2} are separated by depth. \Cseven{C7} shows that also the reflection of $G4$ itself (cf. Figure~\ref{fig:pictures_DAG_CreationProcessEPS}) or a changed property, e.g., node position, did not matter.\\
Interestingly, \emph{edges} and \emph{edge crossings} -- important factors of graph theory and graph aesthetics -- seem not to matter to the participants. The excerpts of \Cthree{C3} and \Cfive{C5} in Figure~\ref{fig:pictures_ClusterExcerpts} clearly show: The edges had no influence on the similarity judgment of the participants. Otherwise DAGs with such different topology would not have been grouped together. The excerpt of \Cseven{C7} shows that  the participants also did not really care about edge crossings. To conclude, we consider the consistency of the hierarchical clusters as high regarding graph theoretical and visual DAG properties. They are also well objectifiable with these properties.

\makeatletter
\newcommand*{\textoverline}[1]{$\overline{\hbox{#1}}\m@th$}
\makeatother

\begin{table}
  \caption{Properties of the DAGs in the clusters \Cone{C1}-\Ceight{C8} along with average easiness (\textoverline{Ease}) and confidence (\textoverline{Conf}) values for each cluster.}
  \label{table:clustersAnalysis}
	\begin{tabularx}{1.0\linewidth}{l>{\centering}X>{\centering}Xp{8.25cm}}
    	\toprule
    	\textbf{Cluster} & \textbf{\textoverline{Ease}} & \textbf{\textoverline{Conf}} & \textbf{DAG Properties} \\
        \midrule
        \Cone{C1} & 4.3 & 4.2 & 
        	 \emph{depth}: 4 
        	 $\bullet$ \emph{number of nodes on level 2}: 4; \emph{on level 3:} 3;\newline \emph{on level 4:} 1
        	 $\bullet$ \emph{leaning}: left\\ 
        \Ctwo{C2} & 4.4 & 4.3 & 
        	 \emph{depth}: 4
        	 $\bullet$ \emph{number of nodes on level 2}: 4; \emph{on level 3:} 3;\newline \emph{on level 4:} 1
        	 $\bullet$ \emph{leaning}: right \\
        \Cthree{C3} & 4.1 & 4.0 & 
        	 \emph{depth}: 3
        	 $\bullet$ \emph{number of nodes on level 2}: 3; \emph{on level 3:} 4  \\
        \Cfour{C4} & 4.1 & 4.1 & 
        	 \emph{depth}: 3
        	 $\bullet$ \emph{number of nodes on level 2}: 3; \emph{on level 3:} 2  \\
        \Cfive{C5} & 3.6 & 3.8 & 
        	 \emph{depth}: 3
        	 $\bullet$ \emph{number of nodes on level 2}: 3; \emph{on level 3:} 3  \\
        \Csix{C6} & 3.7 & 3.7 & 
        	 \emph{depth}: 3
        	 $\bullet$ \emph{number of nodes on level 2}: 4; \emph{on level 3:} 4  \\
        \Cseven{C7} & 3.6 & 3.8 & 
        	 \emph{depth}: 3
        	 $\bullet$ \emph{number of nodes on level 2}: 4; \emph{on level 3:} 3  \\
        \Ceight{C8} & 3.8 & 3.9 & 
        	 \emph{depth}: 3
        	 $\bullet$ \emph{number of nodes on level 2}: 4; \emph{on level 3:} 2 \\
        \bottomrule
    \end{tabularx}
\end{table}	

\subsection{Qualitative Analysis (RQ2)}\label{subsec:qualitative_analysis}
We performed a thematic analysis of the participants' tags to reveal the factors they considered. We also analyzed the factors' importance based on the number of mentions of a specific factor. For this analysis we used the data of all 20 participants since it does not depend on whether a DAG was grouped or not.

\paragraph{Analysis.} First, we transcribed the participants' tags by noting each tag together with how the participant used it, e.g., in a hierarchical manner. Additionally we collected the following data for the tags (henceforth called factors) of each participant: factor type (visual, graph theoretical), combined vs. single factors (e.g., number of levels vs. number of levels \emph{and} number of nodes), number of considered factors, number of values per factor (e.g., number of edge crossings = 1, 2 and 3 $\rightarrow$ number of values = 3). We deemed the factor type as important since the graph theoretical properties are those which contain the information relevant for comparison insights. However, we already know from graph readability research the significant influence of visual factors (e.g., edge crossing). Knowing those for visual comparison is beneficial for controlling their influence. We collected the other data as meta-information on the factors the participants used in order to learn more about the participants' usage of the factors. 

\begin{figure}[t]
  \centering
    \includegraphics[width=\textwidth]{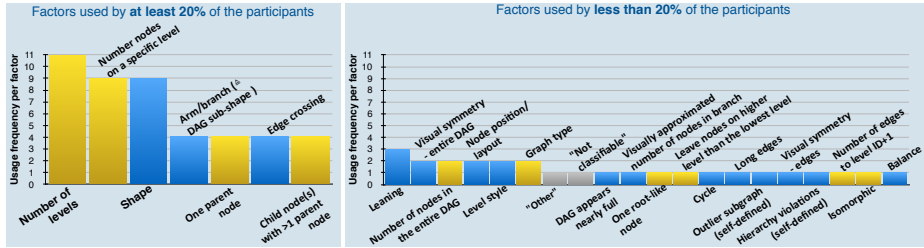}
  \caption{Factors used by the participants (yellow: graph theoretical, blue: visual, gray: no type). Multiple mentions of the same factor by the same participant were excluded.}
  \label{fig:Auswertung_Auswertung_GraphDrawing3}
\end{figure}

\paragraph{Results.} The individual transcriptions can be found on our website. Figure~\ref{fig:Auswertung_Auswertung_GraphDrawing3} shows the factors considered by the participants together with how often a factor was named. Multiple mentions of one and the same factor by one participant were not considered. In total, our participants used 27 distinct factors (cf. Figure~\ref{fig:Auswertung_Auswertung_GraphDrawing3}). Ten of these can be considered to be graph theoretical factors (yellow) and 15 to be visual factors (blue). Two of the used factors are neither graph theoretical nor visual (gray). Just five out of 20 participants used a combined factor and only two of these five used more than one combined factor. The most frequently combined factor was \emph{number of nodes on a specific level} (five times); e.g., number of nodes on the second level = 3 \emph{and} number of nodes on the third level = 4. Eight of the 27 factors were used by at least $20\%$ of the participants (cf. Figure~\ref{fig:Auswertung_Auswertung_GraphDrawing3}, left). We will focus on these eight, for the other 20 factors please refer to Figure~\ref{fig:Auswertung_Auswertung_GraphDrawing3}, right.

The most important factors according to usage frequency were: \emph{number of levels} (i.e., depth of the DAG), \emph{number of nodes on a specific level}, \emph{shape}, \emph{arm/branch ($\equiv$ DAG sub-shape)}, \emph{one parent node}, \emph{edge crossing}, \emph{child node(s) with $>1$ parent node}, \emph{visual leaning (left: \includegraphics[height=0.4cm]{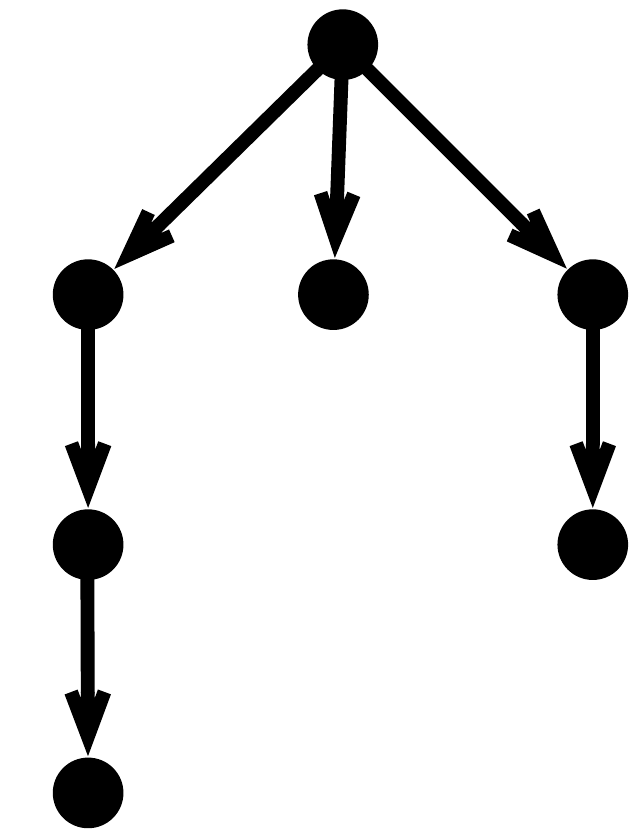}, right: \includegraphics[height=0.4cm]{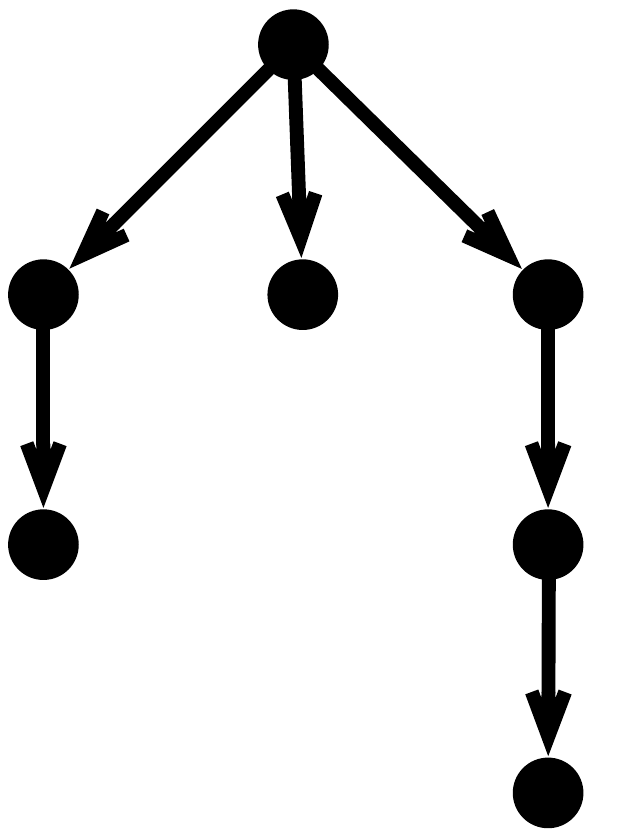})}. The factor \emph{shape} is basically the convex hull of the DAG (\includegraphics[height=0.4cm]{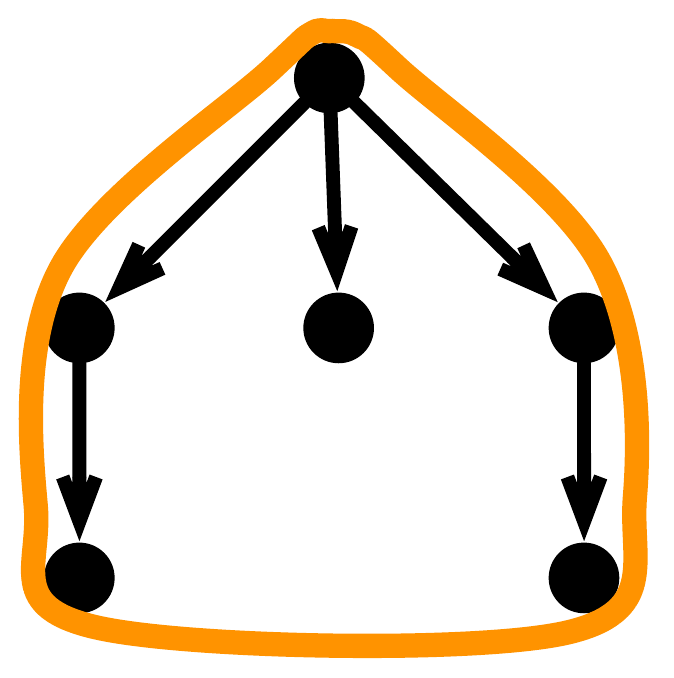}). Regarding \emph{shape} it is interesting to note that we could observe a coherence of \emph{shape} with the \emph{number of nodes on a specific level}. Participants, for instance, denoted a DAG such as \includegraphics[height=0.4cm]{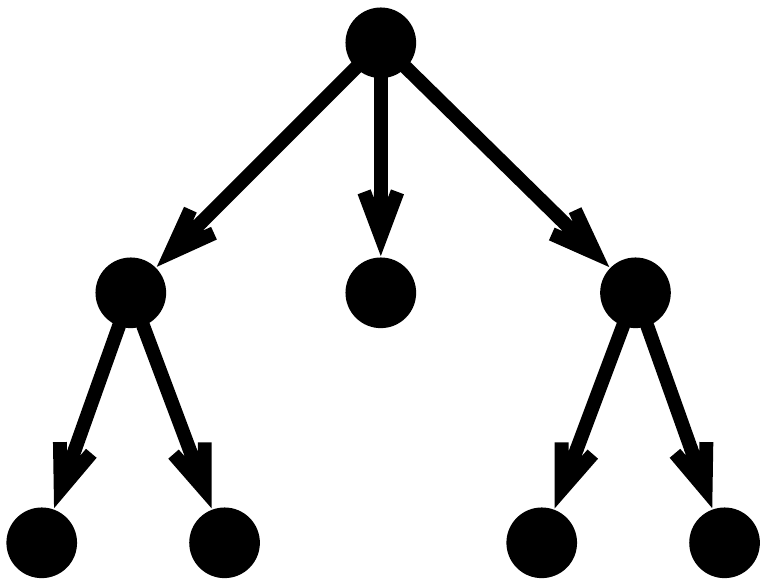} as ``narrow/small pyramid'' and a DAG such as \includegraphics[height=0.4cm]{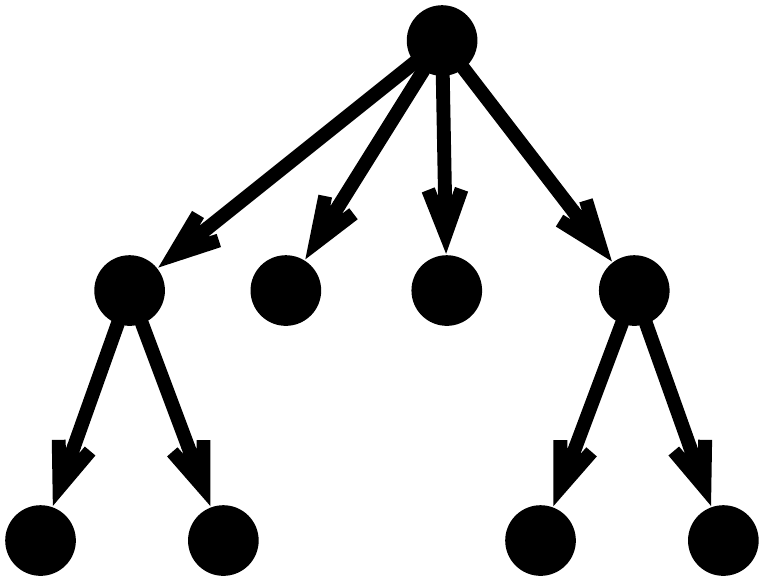} as ``wide/large pyramid''. However, it is clear that this coherence is also influenced by the DAGs' layout. \emph{Arm/branch} refers to the \emph{shape of a DAG's sub-graph} (\includegraphics[height=0.4cm]{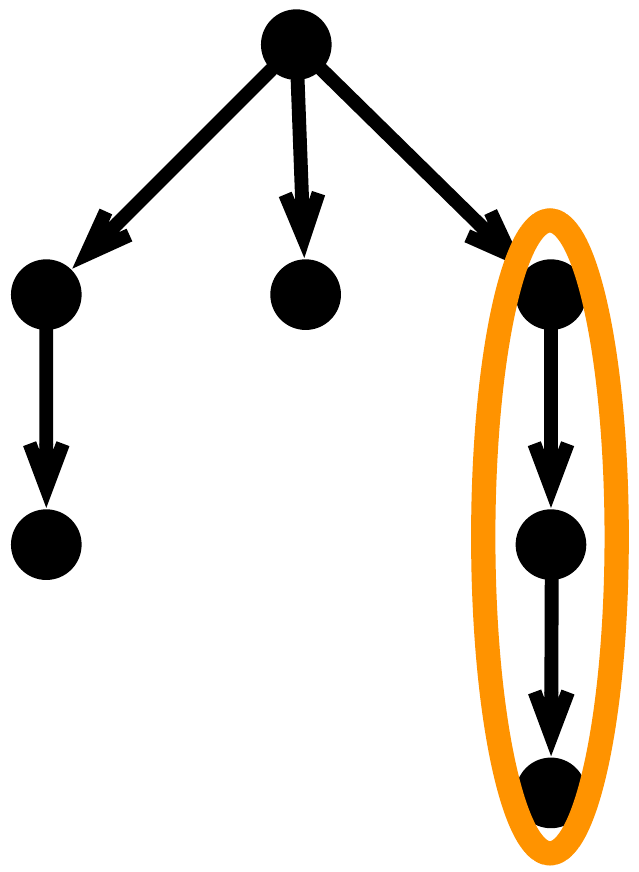}). \emph{Edge crossing} deals with crossings of the visualized edges (\includegraphics[height=0.4cm]{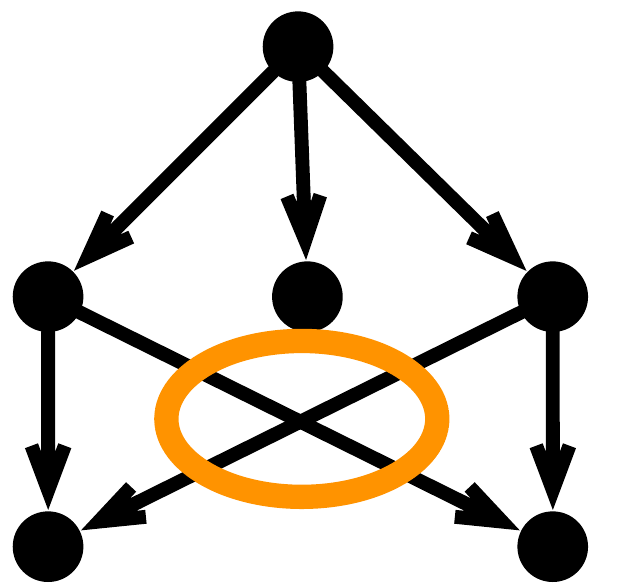}). The participants considered different types of edge crossings, e.g., presence of edge crossing or (un)resolvable edge crossings. The factors \emph{one parent node} and \emph{child node(s) with $>1$ parent node} relate to the number of nodes which are parent to another node (\includegraphics[height=0.4cm]{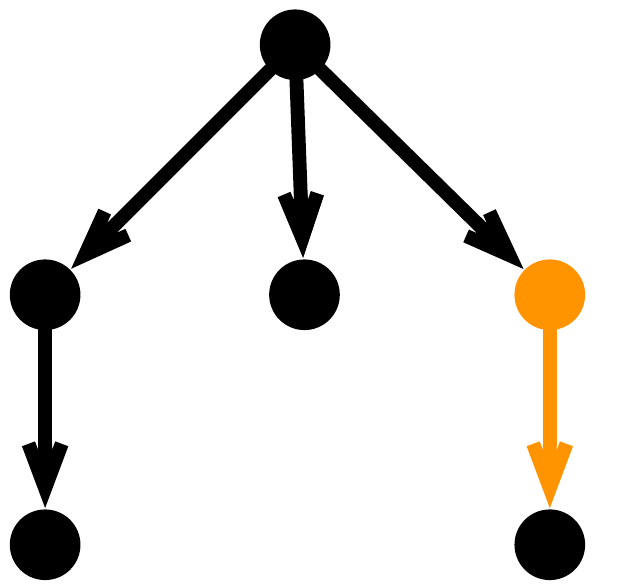}, \includegraphics[height=0.4cm]{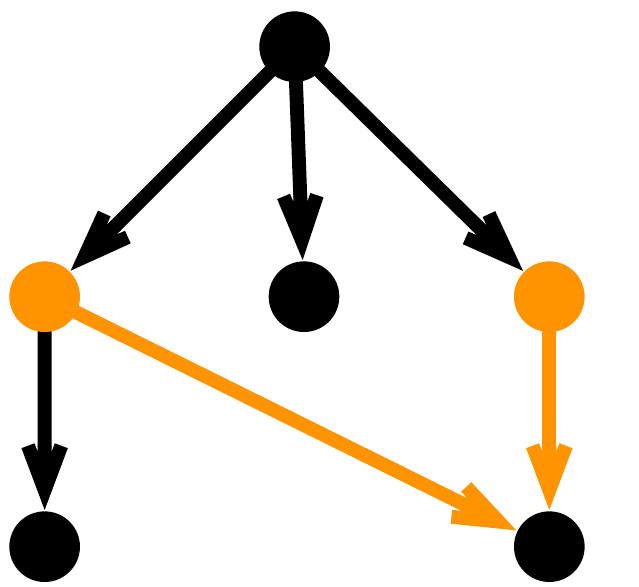}). Again, we could observe that participants used different types of these factors.

Interestingly, also the extracted factors substantiate that edges and edge crossings did not really matter (cf. Section~\ref{subsec:quantitative_analysis}). The factor \emph{edge crossing} is one of the least used of the most important factors. Other edge related factors were used just once (cf. Figure~\ref{fig:Auswertung_Auswertung_GraphDrawing3}, right). Various individual groupings also support this, e.g.: \includegraphics[height=0.4cm]{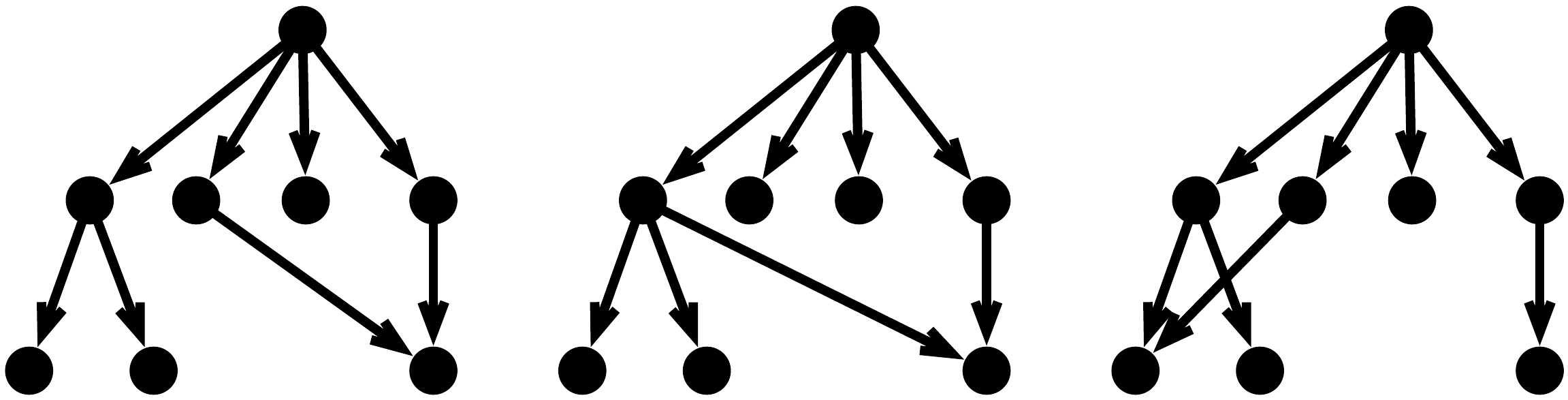} (factor: one parent left).

\section{Discussion and Conclusion}
	\label{sec:discussion}

We conducted a card sorting study to identify the factors influencing the similarity perception of DAGs to mitigate the present knowledge gap regarding this topic despite the vast presence of visual comparison tasks in various disciplines.

Both, the results of our quantitative and qualitative analysis point to \emph{similar} factors which seem to dominantly influence similarity perception of DAGs, namely the \emph{number of levels (depth)}, the \emph{number of nodes on specific levels}, as well as \emph{shape-related aspects} such as the visual leaning of a DAG. Herewith, we can be certain that the self-reported factors of the participants were not biased. The strong influence of shape is remarkable as in our case the spatial arrangement did not convey any additional information. This resulted in cases where structurally identical DAGs were assigned to different groups due to one being left-skewed or right-skewed. Being skewed to the left or right mainly played a role for the 4-level DAGs (cf. \Cone{C1} and \Ctwo{C2}), most likely because it had a stronger influence on the overall shape as in the 3-level cases. Nevertheless, this observation supports previous results which found evidence that perception of graphs is sensitive to its spatial layout (cf. e.g.,~\cite{Huang:2006,mcgrath1997effect}). Surprisingly, edge crossings -- an important factor concerning the readability of graphs~\cite{Purchase:1997a} -- contrary to our expectations did not seem to have a strong impact on perceived DAG similarity. This is, for example, evident in the clusters \Cfive{C5} and \Csix{C6} where no distinction between DAGs with and without edge crossings has been made (cf. Figure~\ref{fig:pictures_ClusterExcerpts}). In the participants' statements we found soft evidence that they did not subconsciously resolve the edge crossing and therefore did not mention edge factors; on the contrary, the edges were not in the focus of the participants.\\
The fixed order of our data items did not lead to arbitrary groupings. The individual groupings and the consensus grouping are well objectifiable with DAG properties. We analyzed the individual groupings by checking the objectifiability of grouped consecutive data items (cf. website for details). The quantitative analysis shows the objectifiability of the consensus grouping (cf. Section~\ref{subsec:quantitative_analysis}).

In future work, it will be necessary to investigate how the identified factors and their importance varies across different graph sizes. It is, for instance, reasonable to assume that, for larger graphs, factors concerning details of a graph (e.g., number of parent nodes, number of nodes on a specific layer) decrease in importance while factors concerning the overall appearance (e.g., shape) increase. Regardless of that, our study provides first results which can contribute to the design of comparative visualizations. Moreover, a better understanding of the factors which drive humans' similarity judgment may also be used towards developing perception-based graph similarity measures. Current notions of graph similarity such as graph isomorphism and edit distance (cf.~\cite{Gao:2010}), descriptive statistics of graph structure measures such as degree distribution or diameter, or iterative approaches which assess the similarity of the neighborhood of nodes (e.g.,~\cite{Jeh:2002,Kleinberg:1999,Melnik:2002}) rely purely on graph theoretical properties.

Besides understanding the individual factors we also deem it important to understand the strategies that participants employ while judging the similarity of data items. This will help to offer useful interactions with comparative visualizations. While our study was not specifically designed for this we could observe circumstantial evidence, as a byproduct from our transcription, that the participants used three distinctive strategies: Eleven participants chose a factor, grouped the entire dataset according to it, and then grouped the resulting groups into further sub-groups (\emph{divide-and-conquer}). There were also seven participants who always \emph{respected the entire dataset considering the factors one after the other}. Some of the participants chose all their factors in advance. Still others chose their factors in an ad hoc fashion; that means, after having grouped the dataset according to a factor they thought about the next. Finally, there were two participants who did their grouping by considering just \emph{one single factor}. More thorough investigations will be necessary to verify these observations.

To conclude, we consider the similarity perception of DAGs in visual comparison across people as consistent and well objectifiable with graph theoretical or visual properties. We find this substantiated by our quantitative and qualitative analysis. An in-depth analysis is subject to future research.

\section{Acknowledgments}
	\label{sec:acknowledgments}
	
This work was financially supported by the Deutsche Forschungsgemeinschaft e.V. (DFG, LA 3001/2-1) and the Austrian Science Fund (FWF, I 2703-N31).

\bibliographystyle{splncs03}
\bibliography{references}

\end{document}